\documentclass[letterpaper,amsfonts,amsmath]{revtex4}
\usepackage[dvips]{epsfig}
\usepackage[dvips]{graphics}
\usepackage{bbold}

\newcommand{\ket}[1]{\ensuremath{\vert{#1}\rangle}}
\newcommand{\bra}[1]{\ensuremath{\langle{#1}\vert}}
\newcommand{\braket}[1]{\ensuremath{\langle{#1}\rangle}}

\begin{document}

\title{Three-State Feshbach Resonances Mediated By Second-Order Couplings}

\author{Christopher J. Hemming}
\email{chemming@chem.ubc.ca}
\affiliation{Department of Chemistry, University of British Columbia,
 2036 Main Mall, Vancouver, British Columbia V6T~1Z1, Canada}

\author{Roman V. Krems}
\email{rkrems@chem.ubc.ca}
\affiliation{Department of Chemistry, University of British Columbia,
 2036 Main Mall, Vancouver, British Columbia V6T~1Z1, Canada}

\date{\today}

\begin{abstract}
We present an analytical study of three-state Feshbach resonances induced by second-order couplings. Such resonances arise when the scattering amplitude is modified by the interaction with a bound state that is not directly coupled to the scattering state containing incoming flux.  Coupling occurs indirectly through an intermediate state. We consider two problems: (i) the intermediate state is a scattering state in a distinct open channel; (ii) the intermediate state is an off-resonant bound state in a distinct closed channel.  The first problem is a model of electric-field-induced resonances in ultracold collisions of alkali metal atoms [Phys.\ Rev.\ A 75, 032709 (2007)]  and the second problem is relevant for ultracold collisions of complex polyatomic molecules, chemical reaction dynamics, photoassociation of ultracold atoms, and electron - molecule scattering. Our analysis yields general expressions for the energy dependence of the T-matrix elements modified by three-state resonances and the dependence of the resonance positions and widths on coupling amplitudes for the weak-coupling limit. We show that the second problem can be generalized to describe resonances induced by indirect coupling through an arbitrary number of sequentially coupled off-resonant bound states and analyze the dependence of the resonance width on the number of the intermediate states.
\end{abstract}

\maketitle

\section{Introduction} 

Following the paper of Tiesinga, Verhaar and Stoof \cite{tiesinga:pra}, Feshbach resonances have been used as an important tool for controlling interactions of ultracold atoms \cite{regal2:prl}, the creation of ultracold molecules \cite{timmermans:physrep,strecker:prl,cubizolles:prl,jochim:prl,jochim:science} and experimental studies of correlated phenomena in ultracold gases \cite{donley:nature,greiner:nature,regal1:prl}. A Feshbach resonance occurs when a scattering state of two colliding particles interacts with a metastable bound state of the two-particle system. The coupling between the scattering state and the bound state leads to a resonant enhancement of the scattering cross section as the energy of the scattering state approaches the energy of the bound state. Collisions at ultracold temperatures are entirely determined by single partial-wave scattering: $s$-wave scattering for  collisions of bosons or distinct atoms and $p$-wave scattering for collisions of identical fermions. Feshbach resonances in ultracold collisions can therefore be described by a two-state model where a single partial-wave state interacts with an isolated bound state (\cite{timmermans:physrep,kohler:rmp} and references in \cite{kohler:rmp}). The energy of the bound state can be tuned with respect to the collision threshold by a magnetic field so magnetic fields have been used in most experiments on Feshbach resonances in ultracold gases (\cite{doyle:eurphysd} and references therein). We have recently proposed that Feshbach resonances in ultracold collisions of distinct atoms may also be tuned by dc electric fields \cite{krems:prl}. An electric field induces couplings between different partial waves of the colliding atoms, which results in a different mechanism of Feshbach resonances.  For example, it is not necessary that the dominant scattering state be directly coupled to the bound state giving rise to resonances so the resonant enhancement of the scattering cross-section may be determined by second order couplings and the resonances cannot be described by a two-state model. The purpose of this paper is to develop a formal theory for three-state Feshbach resonances induced by second-order couplings. 

\begin{figure}
\begin{center}
\scalebox{0.72}{\includegraphics{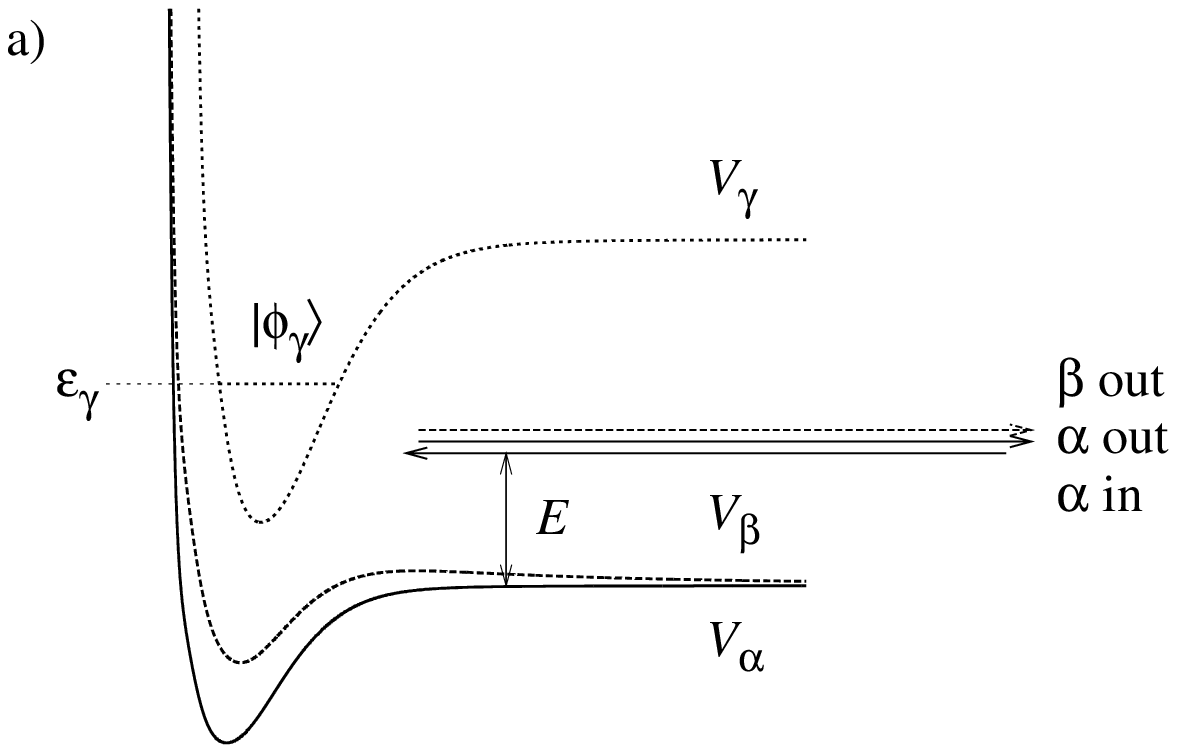}}\\
\vspace*{0.75cm}
\scalebox{0.72}{\includegraphics{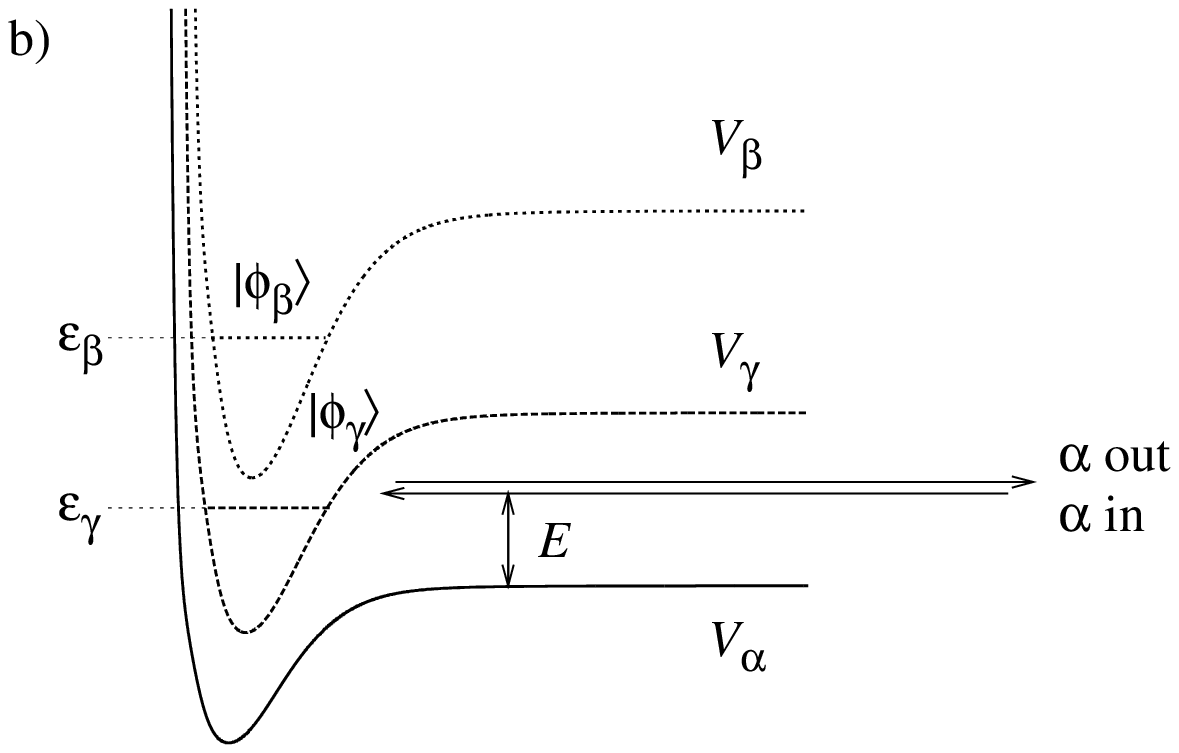}}
\end{center}
\caption{\label{fig}Schematic diagram of a three-state Feshbach resonance in channel $\alpha$ induced by indirect couplings to a bound state in channel $\gamma$ via channel $\beta$: in (a) $\beta$ is a scattering state in an open channel; (b) $\beta$ is a bound state in a closed channel.}
\end{figure}

In the first part of this paper, we use the projection-operator method \cite{feshbach:annphys1, feshbach:annphys2} to analyze three-state Feshbach resonances induced by indirect couplings via an intermediate scattering state.  Figure~\ref{fig}a  illustrates the mechanism of such resonances.  The initial scattering state $\alpha$ is coupled to another scattering state $\beta$, while $\beta$ is coupled to bound state $\gamma$.  The sequence of $\alpha$-$\beta$ and $\beta$-$\gamma$ couplings gives rise to a resonant enhancement in the state $\alpha$. This is a model for electric-field-induced resonances in ultracold collisions of distinct atoms, such as Li and Cs, described in Refs.~[\onlinecite{krems:prl}] and [\onlinecite{likrems:pra}].  Three-state resonances of this type can also be induced by magnetic dipole-dipole interaction \cite{durr:pra}.  D\"{u}rr et al \cite{durr:pra} have recently presented a study of magnetic-dipole-induced predissociation of ultracold Rb$_2$ molecules leading to population of multiple scattering states. 
 
Resonances induced by indirect couplings are also important for collisions of large polyatomic molecules with ultracold atoms.  One proposed method of cooling polyatomic molecules to ultracold temperatures is sympathetic cooling. In the absence of reaction processes, molecules can potentially be cooled by elastic collisions in a reservoir of ultracold atoms \cite{modugno:science,ryjkov:pra,ostendorf:prl}. The experimental realization of this method may, however, be complicated by naturally occurring Feshbach resonances. 
The density of molecular states near zero point energy is very large in polyatomic molecules \cite{nordholm:bbpc} and these bound states may give rise to Feshbach resonances in collisions between molecules and ultracold atoms. The probability of three-body recombination and other loss processes in collisions involving large polyatomic molecules may thus be enhanced. Not all of the bound states  may, however, give rise to Feshbach resonances. The interaction between an atom and a large molecule usually probes only a limited number of molecular states. The other molecular states can interact with the atom - molecule scattering state by indirect couplings and it is not clear whether they can generate Feshbach resonances with significant widths. To elucidate the role of molecular states indirectly coupled to the scattering state, we analyze Feshbach resonances mediated by a sequence of one continuum-bound and several bound-bound couplings. We first consider a three-state Feshbach resonance induced by a combination of two couplings: a coupling between the scattering state and an off-resonant bound state and a coupling between the off-resonant bound state and a resonant bound state (Fig.~\ref{fig}b). In the third part of the paper, we generalize this analysis to a system with Feshbach resonances induced by higher-order couplings.   

Bohn and Julienne examined similar  Feshbach resonances, with second- and higher-order couplings through a series of bound states, in collisions of ultracold atoms induced by optical laser fields  \cite{bohnjulienne:pra}. Their derivation yields the dependence of the scattering amplitude on the Rabi frequency and the detuning of the laser fields.    The problems considered in Sections~\ref{sec:bound} and \ref{sec:multiple} of this paper differ from their systems in that, for cold atom-molecule collisions, the couplings between states depend on the atom-molecule distance and vanish asymptotically.
 
The problems we solve in this paper are equivalent to the diagonalization of a Hamiltonian containing one or more bound states coupled to one or more distinct continua.  Fano presented solutions for the diagonalization of a number of similar Hamiltonians with coupled bound states and continua \cite{fano:pr}.  Our systems differ from the problems solved by Fano in the patterns of couplings present. For example,  Fano considered one bound state coupled directly to multiple continua not coupled to each other,  or a single continuum state coupled to multiple bound states with no couplings between bound states.   Feshbach described a general method for the solution of a continuum coupled to multiple bound states with couplings between states (\cite{feshbach:book}, p.~157); Sections~\ref{sec:bound} and \ref{sec:multiple} represent an application of an approach similar to Feshbach's to the special case of systems of tridiagonal form.

\section{Projection operator methods for Feshbach resonances}
\label{sec:feshbach}
Our approach is based on the projection operator method of Feshbach \cite{feshbach:annphys1,feshbach:annphys2,feshbach:book}.  In this section, we recapitulate the method in order to define the notation.  The Hilbert space of the total Hamiltonian $H$ for the collision system is partitioned into two orthogonal Hilbert spaces:\ ${\mathsf{P}}$, comprising all open channels of the system; and ${\mathsf{Q}}$, containing all closed channels.   These will be referred to as the ${\mathsf{P}}$-channel and the ${\mathsf{Q}}$-channel, although in practice they may contain multiple channels.   

The total wave function of the system, $\ket{\Psi^+(E)}$, belongs to the entire Hilbert space, and satisfies the Schr\"{o}dinger equation 
\begin{equation}
(E-H)\ket{\Psi^+(E)}  = 0\,,\label{eq:genham}
\end{equation}
where $E$ is the total energy of the colliding particles.  The projections of $\ket{\Psi^+}$ onto the ${\mathsf{P}}$ and ${\mathsf{Q}}$ subspaces are given by
\begin{align}
P\ket{\Psi^+} & \equiv \ket{\Psi_P^+}\\
Q\ket{\Psi^+}& \equiv \ket{\Psi_Q}\,,
\end{align}
where $P$ and $Q$ are projection operators.  The projected state vectors satisfy a system of coupled equations
\begin{align}
(E-H_{PP})\ket{\Psi_P^+} & = H_{PQ}\ket{\Psi_Q}\label{eq:Pschrod}\\
(E-H_{QQ})\ket{\Psi_Q} & = H_{QP}\ket{\Psi_P^+}\label{eq:Qschrod}\,,
\end{align}
where $H_{PP}\equiv PHP$, $H_{QQ}\equiv QHQ$, $H_{PQ}\equiv PHQ$, and $H_{QP}\equiv QHP$.  

Formally inverting Eq.~(\ref{eq:Qschrod}) with the Green's operator $(E-H_{QQ})^{-1}$ gives
\begin{equation}
\ket{\Psi_Q} = \frac{1}{E-H_{QQ}}H_{QP}\ket{\Psi_P^+}\,,
\end{equation}
which, when substituted into Eq.~(\ref{eq:Pschrod}), yields 
\begin{equation}
(E-H_{PP})\ket{\Psi_P^+} = H_{PQ}\frac{1}{E-H_{QQ}}H_{QP}\ket{\Psi_P^+}\,.\label{eq:Peff}
\end{equation}
This is an effective Schr\"{o}dinger equation for $\ket{\Psi_P^+}$ 
\begin{equation}
(E-H_{\mathrm{eff}}(E))\ket{\Psi_P^+} = 0\,,
\end{equation}
with an energy-dependent pseudo-Hamiltonian $H_{\mathrm{eff}}(E)=H_{PP}+H_{PQ}(E-H_{QQ})^{-1}H_{QP}$.  The effective potential acting in the ${\mathsf{P}}$-channel, after the effects of the coupling to the ${\mathsf{Q}}$-channel are taken into account, is thus
\begin{equation}
V_{\mathrm{eff}}(E) = V_{PP}+H_{PQ}\frac{1}{E-H_{QQ}}H_{QP}\,,\label{eq:Veff}
\end{equation}
where $V_{PP}$ is the potential energy term of the Hamiltonian $H_{PP}$.  This procedure eliminates $\ket{\Psi_Q}$ from the problem.
 
A scattering state solution for the isolated ${\mathsf{P}}$-channel $\ket{\phi_P^+(E)}$ satisfies
\begin{equation}
(E-H_{PP})\ket{\phi_P^+}  = 0\,,
\end{equation} 
and a free scattering state in the  ${\mathsf{P}}$ subspace $\ket{\chi_P(E)}$ can be obtained from
\begin{equation}
(E-K_{PP})\ket{\chi_P} = 0\,,
\end{equation}
where $K_{PP}$ is the kinetic energy operator of the Hamiltonian $H_{PP}$ such that $H_{PP}=K_{PP}+V_{PP}$.  The functions $\ket{\chi_P}$ and $\ket{\phi_P^+}$ are related by 
\begin{equation}
\ket{\phi_P^+} = \ket{\chi_P}+\frac{1}{E^+-H_{PP}}V_{PP}\ket{\chi_P}\,.\label{eq:phiP+}
\end{equation}
The + superscript on the Green's operator $\frac{1}{E^+-H_{PP}}$ indicates that the calculation is to be performed using the operator $\frac{1}{E+i\eta-H_{PP}}$, with $\eta$ a small positive real number, and the limit of $\eta\to0^+$ should be taken at the end of the calculation.   Eq.~(\ref{eq:phiP+}) gives the scattering state solution for an outgoing wave.  We assume that the system is invariant under time reversal, so the incoming wave scattering state solution corresponding to a free state $\ket{\chi_{P'}}$ is $\ket{\phi_{P'}^{-}}=\ket{\chi_{P'}}+(E^--H_{PP})^{-1}V_{PP}\ket{\chi_{P'}}$.    The minus superscript indicates the operator $\frac{1}{E-i\eta-H_{PP}}$ with the limit $\eta\to0^+$ taken at the end of the calculation.  The subscripts $P$, $P'$ on state vectors denote particular channels within the ${\mathsf{P}}$ subspace.
 
Applying the Green's operator $(E^+-H_{PP})^{-1}$ to Eq.~(\ref{eq:Peff}), we obtain an implicit equation for $\ket{\Psi_P^+}$, 
\begin{equation}
\ket{\Psi_P^+}  = \ket{\phi_P^+} +\frac{1}{E^+-H_{PP}}H_{PQ}\frac{1}{E-H_{QQ}}H_{QP}\ket{\Psi_P^+}\,.\label{eq:psi_implicit}
\end{equation}
With Eqs.~(\ref{eq:phiP+}) and (\ref{eq:psi_implicit}), our choice of the particular free state $\ket{\chi_P}$ selects the incident state and the particular solutions $\ket{\phi_P^+}$ and $\ket{\Psi_P^+}$.  The term $\ket{\phi_P^+}$ appears on the right-hand side of Eq.~(\ref{eq:psi_implicit}) because the general solution of Eq.~(\ref{eq:Peff}) contains a term in the null space of the operator $E-H_{PP}$ which determines the particular solution.  

We are interested in collision-induced transitions to the free state $\ket{\chi_{P'}(E)}$, which in general may not be the same as $\ket{\chi_P(E)}$.   The adjoint of the expression for $\ket{\phi_{P'}^{-}}$ is
\begin{equation}
\bra{\phi_{P'}^{-}} = \bra{\chi_{P'}}+\bra{\chi_{P'}}V_{PP}\frac{1}{E^+-H_{PP}}\,.\label{eq:phiP-}
\end{equation}
The on-shell (i.e.\ between states of the same energy) T-matrix element $T_{P'P}(E)$ is the probability amplitude for a transition from the incoming free state $\ket{\chi_P(E)}$ to the outgoing free state $\ket{\chi_{P'}(E)}$.  It is given by 
\begin{align}
T_{P'P} 
  &= \braket{\chi_{P'}|V_{\mathrm{eff}}|\Psi_P^+}\nonumber\\
   & =  \braket{\chi_{P'}|V_{PP}|\phi_P^+} + \braket{\phi_{P'}^{-}|H_{PQ}\frac{1}{E-H_{QQ}}H_{QP}|\Psi_P^+}\label{eq:Tb_Tr}\\
   & \equiv  T^b_{P'P} + T_{P'P}^r \nonumber\,,
\end{align}
where $T^b_{P'P}$ is the transition amplitude for background scattering in the isolated ${\mathsf{P}}$-channel and $T^r_{P'P}$ is the resonant contribution due to coupling to the ${\mathsf{Q}}$-channel \cite{taylor:book}. Eqs.~(\ref{eq:psi_implicit}) and (\ref{eq:phiP-}) have been used to obtain the second line of Eq.~(\ref{eq:Tb_Tr}).    
We assume that the ${\mathsf{Q}}$-channel contains an isolated bound state $\ket{\phi_1}$ giving rise to the resonance, so that
\begin{equation}
\frac{1}{E-H_{QQ}} \approx  \frac{\ket{\phi_1}\bra{\phi_1}}{E-\epsilon_1}\,,
\end{equation}
where $\epsilon_1$ is the energy of the bound state.
Using this expression, we can write for $T^r_{P'P}$,
\begin{equation}
T^r_{P'P} = \frac{\braket{\phi_{P'}^{-}|H_{PQ}|\phi_1}\braket{\phi_1|H_{QP}|\Psi_P^+}}{E-\epsilon_1}\,.\label{eq:Tr}
\end{equation}
To evaluate the matrix element $\braket{\phi_1|H_{QP}|\Psi_P^+}$ we multipy Eq.~(\ref{eq:psi_implicit}) on the left by $\bra{\phi_1}H_{QP}$ and solve to obtain
\begin{equation}
\braket{\phi_1|H_{QP}|\Psi_P^+} = \frac{\braket{\phi_1|H_{QP}|\phi_P^+}}{1-\cfrac{\braket{\phi_1|H_{QP}\frac{1}{E^+-H_{PP}}H_{PQ}|\phi_1}}{E-\epsilon_1}}\,.
\end{equation}
Substituting this into Eq.~(\ref{eq:Tr}) we obtain
\begin{equation}
T^r_{P'P} =  \frac{ \braket{\phi_{P'}^{-}|H_{PQ}|\phi_Q}  \braket{\phi_Q|H_{QP}|\phi_P^+} }{E-\epsilon_1- \braket{\phi_Q|H_{QP}\frac{1}{E^+-H_{PP}}H_{PQ}|\phi_Q} }\,.\label{eq:Tr0}
\end{equation}
The Green's operator in the denominator of this equation may be expanded in the basis $\ket{\phi_{P''}^+(E')}$,
\begin{equation}
\frac{1}{E^+-H_{PP}} = \int_0^\infty dE' \sum_{P''}\frac{\ket{\phi_{P''}^+(E')}\bra{\phi_{P''}^+(E')}}{E^+-E'}\,,\label{eq:Gppexpansion}
\end{equation}
which gives
\begin{equation}
{\mathrm{Re}}\,\braket{\phi_Q|H_{QP}\frac{1}{E^+-H_{PP}}H_{PQ}|\phi_Q} = {\mathcal{P}}\int_0^\infty dE' \frac{|\braket{\phi_Q|H_{QP}|\phi_P^+(E')}|^2}{E-E'}\equiv\Delta(E)\,,
\label{eq:Deltadefn}
\end{equation}
where $\mathcal{P}$ indicates the Cauchy principal value integral, in which the symmetric limit approaching the singularity at $E'=E$ is taken.  Note that, as a result of taking the real part of the expression, the complex-valued $E^+$ in the energy denominator of Eq.~(\ref{eq:Gppexpansion}) is replaced by the real-valued $E$.

Since
\begin{equation}
\lim_{\eta\to0^+} \frac{1}{E-i\eta-E'}-\frac{1}{E+i\eta-E'} = 2\pi i\delta(E-E')\,,\label{eq:deltafunction}
\end{equation}
we obtain
\begin{equation}
{\mathrm{Im}}\,\braket{\phi_Q|H_{QP}\frac{1}{E^+-H_{PP}}H_{PQ}|\phi_Q} = -\pi \sum_{P''} |\braket{\phi_Q|H_{QP}|\phi_{P''}^+(E)}|^2 \equiv -\frac{\Gamma}{2}\,.
\label{eq:Gammadefn}
\end{equation}
With the additional definition
\begin{equation}
\Gamma_{P'P} \equiv 2\pi\braket{\phi_{P'}^{-}|H_{PQ}|\phi_Q}  \braket{\phi_Q|H_{QP}|\phi_P^+}\,,\label{eq:GammaPPdefn}
\end{equation}
we thus obtain from Eq.~(\ref{eq:Tr0})
\begin{equation}
T_{P'P}^r = \frac{\Gamma_{P'P}(E)}{2\pi\left(E-\epsilon_1-\Delta(E)+\frac{i\Gamma(E)}{2}\right)}\,.\label{eq:Tr0a}
\end{equation}

The $S$ operator is related to the $T$ operator by 
\begin{equation}
S = \mathbb{1}-2\pi i T\,.
\end{equation}
The matrix elements of $S$ in the ($P$, $E$) basis are
\begin{equation}
\braket{\chi_{P'}(E')|S|\chi_P(E)} = \delta(E-E')s_{P'P}(E) = \delta(E-E')(\delta_{P'P}-2\pi i T_{P'P}(E))\,.\label{eq:defslittles}
\end{equation}
$S$ may be expressed as the sum of background and resonant parts:\ $S= S^b+S^r$, and similarly $s_{P'P}=s^b_{P'P}+s^r_{P'P}$.  The background $S$ operator $S^b = \mathbb{1}-2\pi iT^b$ describes scattering in the open channels in the absence of coupling to the closed channels. The resonant part of $s_{P'P}$ is 
\begin{equation}
s^r_{P'P} = \frac{-i\Gamma_{P'P}(E)}{E-\epsilon_1-\Delta(E)+\frac{i\Gamma(E)}{2}}\,.\label{eq:Selement}
\end{equation}
Using Eq.~(\ref{eq:phiP-}), the adjoint of Eq.~(\ref{eq:phiP+}), and Eq.~(\ref{eq:deltafunction}), one may show that
\begin{equation}
\bra{\phi_{P'}^-} = \sum_{P''}\left(\delta_{P'P''}-2\pi i\braket{\chi_{P'}|V_{PP}|\phi_{P''}^+}\right)\bra{\phi_{P''}}\,.\label{eq:factoring1} 
\end{equation}
The quantity in parentheses is $s^b_{P'P''}$,  hence from Eq.~(\ref{eq:GammaPPdefn}) we find
\begin{equation}
\Gamma_{P'P} = \sum_{P''} 2\pi s^b_{P'P''}\braket{\phi_{P''}^+|H_{PQ}|\phi_Q}\braket{\phi_Q|H_{QP}|\phi_P^+}\,.\label{eq:factoring2}
\end{equation}
Since $S^b$ is unitary,
\begin{equation}
|\Gamma_{PP}| = 2\pi|\braket{\phi_Q|H_{QP}|\phi_P^+}|^2
\end{equation}
and $\displaystyle\Gamma=\sum_{P}|\Gamma_{PP}|$.  Eq.~(\ref{eq:Selement}) shows that the resonance gives rise to a peak in all open channels, with the magnitude of the resonant scattering in channel $P$ determined by $|\Gamma_{PP}|$.  The position and width of the resonance are the same in all channels, and the position $E_r$ satisfies the equation $E_r-\epsilon_1-\Delta(E_r)=0$.  If the resonance is sufficiently narrow and $\Gamma$ does not vary significantly, then its width is approximately $\Gamma(E_r)$. 

If the open channel space consists of a single partial wave, the single element $s_{PP}$ may be written as $s_{P}(E) = e^{2i\delta}$, where $\delta(E)$ is the scattering phase shift.   The background part of $s_P$ is $s^b_P=e^{2i\delta_b}$,  and $\delta=\delta_b+\delta_r$.   Eq.~(\ref{eq:factoring2}) takes the form $2\pi\braket{\phi_P^-|H_{PQ}|\phi_Q}\braket{\phi_Q|H_{QP}|\phi_P^+}=e^{2i\delta_b}\Gamma$.  Using Eqs.~(\ref{eq:GammaPPdefn}) and (\ref{eq:Tr0a}), we obtain
\begin{equation}
s_{P}(E) = e^{2i\delta_b}\left(\frac{E-\epsilon_1-\Delta-\frac{i\Gamma}{2}}{E-\epsilon_1-\Delta+\frac{i\Gamma}{2}}\right)\,,
\end{equation}
and find that the resonant phase shift is given by 
\begin{equation}
\delta_r(E) = \tan^{-1} \frac{-\Gamma(E)}{2(E-\epsilon_1-\Delta(E))}\,.
\end{equation}

\section{Intermediate scattering state}
\label{sec:scattering}

In this section, we apply the formalism of Sec.~\ref{sec:feshbach} to a system with three channels:\ channel $\alpha$, channel $\beta$, and channel $\gamma$.  The state $\alpha$ is an incoming channel for the colliding particles, the state $\beta$ is a distinct open channel, and the state $\gamma$ is a closed channel.  There is no direct coupling between $\alpha$ and $\gamma$ but the incoming collision channel is coupled to $\gamma$ indirectly via a sequence of two couplings: $V_{\alpha\beta}$ and $V_{\beta\gamma}$.  In our motivating example of a collision between two atoms of different types \cite{krems:prl,likrems:pra}, $\alpha$ is the $s$-wave scattering state of the ground electronic state, $\beta$ is the $p$-wave scattering state of the ground state, and $\gamma$ is a $p$-wave bound state of an excited electronic state.  Figure~\ref{fig}a illustrates this system.  We denote projection operators onto the three channels by $P_\alpha$, $P_\beta$, $P_\gamma$, and the projections $\ket{\alpha^+}=P_\alpha\ket{\Psi^+}$, $\ket{\beta^+}=P_\beta\ket{\Psi^+}$, and $\ket{\gamma}=P_\gamma\ket{\Psi^+}$.  We also define projected Hamiltonians  $H_{\nu\rho}=P_\nu H P_\rho$, where $\nu$, $\rho=\alpha$, $\beta$ or $\gamma$.  

The coupled equations for our system are (recalling that $V_{\alpha\gamma}=V_{\gamma\alpha}=0$)
\begin{align}
(E-H_{\alpha\alpha})\ket{\alpha^+} & = V_{\alpha\beta}\ket{\beta+}\label{eq:alphachannel1}\\
(E-H_{\beta\beta})\ket{\beta^+} & = V_{\beta\alpha}\ket{\alpha^+}+V_{\beta\gamma}\ket{\gamma}\label{eq:betachannel1}\\
(E-H_{\gamma\gamma})\ket{\gamma} & = V_{\gamma\beta}\ket{\beta^+}\label{eq:gammachannel1}\,.
\end{align}
Because $H$ is Hermitian, $V_{\alpha\beta}=V_{\beta\alpha}^\dag$ and $V_{\beta\gamma}=V_{\gamma\beta}^\dag$.

In our example of atom-atom collisions, the projection operators $P_\nu$ ($\nu=\alpha$,$\beta$,$\gamma$) are of the form $\ket{n\ell m_\ell}\bra{n\ell m_\ell}$, where $n$ denotes the electronic states of the colliding atoms and $\ell$ and $m_\ell$ are nuclear angular momentum quantum numbers.  The channel Hamiltonians $H_{\nu\nu}$ are the nuclear radial Hamiltonians and the states $\ket{\alpha^+}$, $\ket{\beta^+}$, $\ket{\gamma}$ correspond to functions of the internuclear radial coordinate $R$:\ $\psi_\alpha^+(R)=\braket{R|\alpha+}$,  $\psi_\beta^+(R)=\braket{R|\beta+}$, $\psi_\gamma(R)=\braket{R|\gamma}$.  The centrifugal potential $\ell_\nu(\ell_\nu+1)/2\mu R^2$ is retained as part of the kinetic energy operator $K_{\nu\nu}$.   The states $\ket{\alpha^+}$ and $\ket{\beta^+}$  carry the + sign because they are nonnormalizable scattering states with proper asymptotic forms as $R\to\infty$.  The state $\ket{\gamma}$ is a normalizable bound state, which means that $R\psi_\gamma(R)$ decays exponentially as $R\to\infty$.   The off-diagonal interaction potentials $V_{\alpha\beta}(R)=\braket{R|V_{\alpha\beta}|R}$ and $V_{\beta\gamma}(R)=\braket{R|V_{\beta\gamma}|R}$ are functions of $R$.    Our goal is to find the dependence of scattering properties on the coupling strength, so we assume that $V_{\alpha\beta}(R)=\Omega_1\tilde{V}_{\alpha\beta}(R)$ and $V_{\beta\gamma(R)}=\Omega_2\tilde{V}_{\beta\gamma}(R)$, where $\Omega_1$ and $\Omega_2$ represent  the coupling intensities and $\tilde{V}_{\alpha\beta}(R)$ and $\tilde{V}_{\beta\gamma}(R)$ are functions independent of the coupling intensities.  In our example of Li-Cs collision in electric fields, $\tilde{V}_{\alpha\beta}$ can be approximated by a Gaussian function, whereas $\Omega_1$ is proportional to the strength of the electric field \cite{likrems:pra}.

The ${\mathsf{P}}$ subspace defined in Section~\ref{sec:feshbach} comprises two open channels, $\alpha$ and $\beta$, and the closed channel subspace ${\mathsf{Q}}$ is the $\gamma$ channel:\ 
\begin{align}
H_{PP} & = \left( \begin{array}{cc}H_{\alpha\alpha} & V_{\alpha\beta} \\ V_{\beta\alpha} & H_{\beta\beta} \end{array}\right)\label{eq:Hpp}\\
H_{PQ} & = \left( \begin{array}{c} 0 \\ V_{\beta\gamma} \end{array} \right)\\
H_{QP} & = \left( \begin{array}{cc} 0 & V_{\gamma\beta} \end{array} \right)\\
H_{QQ} & = \left( H_{\gamma\gamma} \right)\,.
\end{align}
 
The final solution can be described in terms of Green's operators containing various parts of the Hamiltonian.  The following section provides useful notation and identities for the Green's operators for this problem.

\subsection{Green's operator notation and identities}
\label{sec:notation}

Here the symbols $\nu$ and $\rho$ are used to denote any of $\alpha$, $\beta$ or $\gamma$.  We denote the  Green's operator for the outgoing wave in the isolated $\nu$-channel by
\begin{equation}
G_\nu^+(E) \equiv \frac{1}{E^+-H_{\nu\nu}}\,.
\end{equation}
This operator satisfies the equation $G_\nu^+(E)(E-H_{\nu\nu})={\mathbb{1}}_\nu$, where ${\mathbb{1}}_\nu$ is the identity operator in the Hilbert space of channel $\nu$.  Similarly, $G_\nu^-(E)$ represents the incoming wave Green's operator,  while $G_\nu(E)$ is a Green's operator with neither + nor - specified.  It is used to describe closed channels, where $G_\nu^+(E)=G_\nu^-(E)$.

We make use of Green's operators of the form 
\begin{equation}
\mathcal{G}^+_{\nu(\rho+)}(E) \equiv \frac{1}{E^+-H_{\nu\nu}-V_{\nu\rho}G^+_\rho(E)V_{\rho\nu}}\,,
\end{equation}
where the + superscript on the symbol $\mathcal{G}^+_{\nu(\rho+)}$ indicates the presence of a + superscript on the $E$ in the denominator, and the + following the index $\rho$ in parentheses indicates the + superscript in the $V_{\nu\rho}G^+_\rho V_{\rho\nu}$ term.  We will use similar symbols where either or both of these superscripts may be replaced with a minus sign or with no superscript.  The argument to a Green's operator is assumed to be $E$, unless specified otherwise.

A useful identity is
\begin{equation}
\mathcal{G}^+_{\nu(\rho+)}V_{\nu\rho}G^+_\rho = G_{\nu}^+V_{\nu\rho}\mathcal{G}^+_{\rho(\nu+)}\,,\label{eq:identity1}
\end{equation}
which we derive starting from
\begin{equation}
V_{\nu\rho}({\mathbb{1}}_\rho-G^+_\rho V_{\rho\nu} G^+_\nu V_{\nu\rho}) = ( {\mathbb{1}}_\nu - V_{\nu\rho}G_\rho^+V_{\rho\nu} G_\nu^+ )V_{\nu\rho}\,.
\end{equation}
Writing ${\mathbb{1}}_\rho=G_\rho^+(E-H_{\rho\rho})$ and ${\mathbb{1}}_{\nu}=(E-H_{\nu\nu})G_\nu^+$, we obtain after factorization
\begin{equation}
V_{\nu\rho}G_\rho^+(E-H_{\rho\rho}-V_{\rho\nu}G_\nu^+V_{\nu\rho}) = (E-H_{\nu\nu}-V_{\nu\rho}G_\rho^+V_{\rho\nu})G_{\nu}^+V_{\nu\rho}\,,
\end{equation}
which yields Eq.~(\ref{eq:identity1}) after multiplying on the left by ${\mathcal{G}}_{\nu(\rho+)}^+$ and on the right by ${\mathcal{G}}_{\rho(\nu+)}^+$.

Another useful identity is
\begin{equation}
\mathcal{G}_{\nu(\rho)}^+ = G_{\nu}^+ + G_\nu^+V_{\nu\rho}G_\rho^+V_{\rho\nu}\mathcal{G}_{\nu(\rho)}^+\,,\label{eq:identity2}
\end{equation}
which is the general linear operator identity $\frac{1}{A-B}=\frac{1}{A}+\frac{1}{A}B\frac{1}{A-B}$ applied to ${\mathcal{G}}_{\nu(\rho+)}^+$.

\subsection{The Green's operator $\frac{1}{E^+-H_{PP}}$}
\label{sec:greens}

To apply the method of Section~\ref{sec:feshbach} we must evaluate the Green's operator $\frac{1}{E^+-H_{PP}}$,   which enters Eqs.~(\ref{eq:phiP+}) and (\ref{eq:psi_implicit}).
We do this by solving the inhomogeneous Schr\"{o}dinger equation
\begin{equation}
(E-H_{PP})\ket{b} = \ket{c}\,,\label{eq:inhomog}
\end{equation}
where 
\begin{equation}
\ket{c} = \left( \begin{array}{c} \ket{c_\alpha} \\ \ket{c_\beta} \end{array}\right)
\end{equation}
is a given arbitrary vector with no components on the null space of $E-H_{PP}$, which is necessary for a solution to Eq.~(\ref{eq:inhomog}) to exist.   Every solution to Eq.~(\ref{eq:inhomog}) corresponding to the correct boundary conditions is of the form
\begin{equation}
\ket{b} =  \ket{b'}+\frac{1}{E^+-H_{PP}}\ket{c}\,,\label{eq:inhomoggensol}
\end{equation}
where $\ket{b'}$ is a solution to the homogeneous equation.  Using the identity
\begin{equation}
\ket{b} = \left( \begin{array}{c} \ket{b_\alpha} \\ \ket{b_\beta} \end{array}\right)\,,
\end{equation}
Eq.~(\ref{eq:inhomog}) can be rewritten as
\begin{align}
(E-H_{\alpha\alpha})\ket{b_\alpha} & = V_{\alpha\beta}\ket{b_\beta} + \ket{c_\alpha}\label{eq:inhomogA}\\
(E-H_{\beta\beta})\ket{b_\beta} & =  V_{\beta\alpha}\ket{b_\alpha} + \ket{c_\beta}\label{eq:inhomogB}\,.
\end{align}
Applying, respectively, $G_\alpha^+$ and $G_\beta^+$ to these equations we obtain
\begin{align}
\ket{b_\alpha} & = \ket{b_\alpha'} + G_\alpha^+V_{\alpha\beta}\ket{b_\beta} + G_\alpha^+\ket{c_\alpha}\\
\ket{b_\beta} & = \ket{b_\beta'} +G_\beta^+V_{\beta\alpha}\ket{b_\alpha} +G_\beta^+\ket{c_\beta}\,,
\end{align}
where $\ket{b'_\alpha}$ and $\ket{b'_\beta}$ satisfy the equations $(E-H_{\alpha\alpha})\ket{b_\alpha'}=0$ and $(E-H_{\beta\beta})\ket{b_\beta'}=0$.  Substituting these equations into the right-hand sides of Eqs.~(\ref{eq:inhomogA}) and (\ref{eq:inhomogB}) and rearranging we obtain
\begin{align}
(E-H_{\alpha\alpha} -V_{\alpha\beta}G_\beta^+V_{\beta\alpha})\ket{b_\alpha} & = V_{\alpha\beta}\ket{b_\beta'} +V_{\alpha\beta}G_\beta^+\ket{c_\beta'} + \ket{c_\alpha}\label{eq:effective_ba}\\
(E-H_{\beta\beta} -V_{\beta\alpha}G_\alpha^+V_{\alpha\beta})\ket{b_\beta} & = V_{\beta\alpha}\ket{b_\alpha'}+V_{\beta\alpha}G_\alpha^+\ket{c_\alpha'}+\ket{c_\beta}\,.\label{eq:effective_bb}
\end{align}
Multiplying Eqs.~(\ref{eq:effective_ba}) and (\ref{eq:effective_bb}) by $\mathcal{G}_{\alpha(\beta+)}^+$ and $\mathcal{G}_{\beta(\alpha+)}^+$ respectively, and rewriting the system in matrix form, we obtain
\begin{equation}
\left(\begin{array}{c} \ket{b_\alpha} \\ \ket{b_\beta} \end{array}\right) = 
\left( \begin{array}{cc} 
\mathcal{G}_{\alpha(\beta+)}^+  & \mathcal{G}_{\alpha(\beta+)}^+V_{\alpha\beta}G_{\beta}^+ \\
 \mathcal{G}^+_{\beta(\alpha+)}V_{\beta\alpha}G_\alpha^+ & \mathcal{G}^+_{\beta(\alpha+)}
 \end{array} \right)
 \left( \begin{array}{c}  \ket{c_\alpha} \\ \ket{c_\beta}
 \end{array}\right) + 
 \left( \begin{array}{c}
\ket{b_\alpha''}+{\mathcal{G}}_{\alpha(\beta+)}^+V_{\alpha\beta}\ket{b_\beta'}
\\
\ket{b_\beta''}+\mathcal{G}_{\beta(\alpha+)}^+V_{\beta\alpha}\ket{b_\alpha'}
\end{array} 
 \right)\,,
\end{equation}
where $(E-H_{\alpha\alpha}-V_{\alpha\beta}G_\beta^+V_{\beta\alpha})\ket{b_\alpha''}=0$ and $(E-H_{\beta\beta}-V_{\beta\alpha}G_\alpha^+V_{\alpha\beta})\ket{b_\beta''}=0$.  This is of the form of Eq.~(\ref{eq:inhomoggensol}) -- the reader may verify that the last term on the right-hand side is a solution to the homogeneous equation -- and we conclude that
\begin{equation}
\frac{1}{E^+-H_{PP}} = \left( \begin{array}{cc} 
\mathcal{G}_{\alpha(\beta+)}^+(E)  & \mathcal{G}_{\alpha(\beta+)}^+(E)V_{\alpha\beta}G_{\beta}^+(E) \\
 \mathcal{G}^+_{\beta(\alpha+)}(E)V_{\beta\alpha}G_\alpha^+(E) & \mathcal{G}^+_{\beta(\alpha+)}(E)
 \end{array} \right)\,,\label{eq:greens}
\end{equation}
which can be verified by matrix multiplication with $E-H_{PP}$.

\subsection{Background scattering in the open channel}

Before including the coupling to the ${\mathsf{Q}}$-channel, it is necessary to obtain the background scattering properties in the ${\mathsf{P}}$-channel.  These solutions are written in  terms of scattering solutions for the isolated $\alpha$- and $\beta$- channels.   A free scattering state for the $\alpha$-channel $\ket{\alpha_0(E)}$ satisfies the equation $(E-K_{\alpha\alpha})\ket{\alpha_0}=0$.  Similarly, a $\beta$-channel free state $\ket{\beta_0(E)}$ satisfies $(E-K_{\beta\beta})\ket{\beta_0}=0$. In the example of atom-atom collisions with the $\alpha$ and $\beta$ channels corresponding to $s$ and $p$ partial waves, $\braket{R|\alpha_0(E)}=\hat{\jmath}_0((2\mu E/\hbar^2)^{1/2}R)$ and $\braket{R|\beta_0(E)}=\hat{\jmath}_1((2\mu E/\hbar^2)^{1/2}R)$, where $\hat{\jmath}_\ell$ are Riccati-Bessel functions.  Eq.~(\ref{eq:phiP+}) for the isolated $\alpha$ and $\beta$ channels gives
\begin{align}
\ket{\phi_\alpha^+} & = \ket{\alpha_0}+G_{\alpha}^+V_{\alpha\alpha} \ket{\alpha_0}\label{eq:phiA+andA0}\\
\ket{\phi_\beta^+} &= \ket{\beta_0}+G_{\beta}^+ V_{\beta\beta} \ket{\beta_0}\,,\label{eq:phiB+andB0}
\end{align}
where $\ket{\phi_\alpha^+(E)}$ is a scattering state solution for the isolated $\alpha$ channel, satisfying the equation $(E-H_{\alpha\alpha})\ket{\phi_\alpha^+(E)}=0$, and $\ket{\phi_\beta^+(E)}$ is a scattering state for the isolated $\beta$-channel and $(E-H_{\beta\beta})\ket{\phi_\beta^+(E)}=0$.

For the ${\mathsf{P}}$ channel background scattering problem, we seek states $\ket{\phi_P^+}$ satisfying $(E-H_{PP})\ket{\phi_P+}=0$ with appropriate boundary conditions.   
A basis for free states of the two-dimensional ${\mathsf{P}}$ channel is given by the vectors
\begin{align}
\ket{\chi_P^\alpha} & = \left(\begin{array}{c} \ket{\alpha_0} \\ 0 \end{array}\right)\label{eq:chiA}\\ 
\ket{\chi_P^\beta} & =\left(\begin{array}{c}  0 \\ \ket{\beta_0}  \end{array}\right)\label{eq:chiB}\,.
\end{align}
Equation~(\ref{eq:phiP+}) relates the scattering state $\ket{\phi_P^+}$ to a free state $\ket{\chi_P}$ in the space of the ${\mathsf{P}}$ channel.  Using Eq.~(\ref{eq:greens}) for the Green's operator $\frac{1}{E^+-H_{PP}}$, as well as Eqs.~(\ref{eq:Hpp}), (\ref{eq:identity1}), (\ref{eq:identity2}),  (\ref{eq:phiA+andA0}), and (\ref{eq:phiB+andB0}), we solve Eq.~(\ref{eq:phiP+}) for the free states (\ref{eq:chiA}) and (\ref{eq:chiB}) to obtain
\begin{align}
\ket{\phi_P^{\alpha+}} 
= 
\left( \begin{array}{c} \ket{\alpha^+_{\mathrm{BG}}} \\ \ket{\beta^+_{\mathrm{BG}}} \end{array}\right) 
& = 
\left( \begin{array}{c} 
\ket{\phi_\alpha^+} + \mathcal{G}_{\alpha(\beta+)}^+V_{\alpha\beta}G^+_{\beta}V_{\beta\alpha}\ket{\phi_\alpha^+} \\
\mathcal{G}^+_{\beta(\alpha+)}V_{\beta\alpha}\ket{\phi_\alpha^+}
\end{array}\right)\label{eq:phiPA+2}
\\  
\ket{\phi_P^{\beta+}} =  
\left( \begin{array}{c}  \ket{\alpha^{\prime+}_{\mathrm{BG}}} \\ \ket{\beta^{\prime+}_{\mathrm{BG}}} \end{array} \right) 
& =  \left(  
\begin{array}{c}
\mathcal{G}^+_{\alpha(\beta+)}V_{\alpha\beta}\ket{\phi_\beta^+} 
\\ 
\ket{\phi_\beta^+} + \mathcal{G}_{\beta(\alpha+)}^+V_{\beta\alpha}G^+_{\alpha}V_{\alpha\beta}\ket{\phi_\beta^+}
\end{array}
\right)\,.\label{eq:phiPB+2}
\end{align}
The first solution, $\ket{\phi_P^\alpha+}$, represents a state with incoming flux only in the $\alpha$ channel, while the state $\ket{\phi_P^{\beta+}}$ corresponds to incoming flux in the $\beta$ channel only.  

For elastic scattering in the $\alpha$-channel, the background part of the on-shell T-matrix element, from Eq.~(\ref{eq:Tb_Tr}), is
\begin{align}
T_{\alpha\alpha}^b & = \braket{\chi_P^\alpha|V_{PP}|\phi^{\alpha+}_P}\nonumber\\
& = \braket{\alpha_0|V_{\alpha\alpha}|\alpha_{\mathrm{BG}}^+} + \braket{\alpha_0|V_{\alpha\beta}|\beta_{\mathrm{BG}}^+}\,.
\end{align} 

\subsection{Transition amplitudes}

The effective potential in the P-channel, defined by Eq.~(\ref{eq:Veff}), takes the form
\begin{equation}
V_{\mathrm{eff}}(E)  = \left( \begin{array}{cc} V_{\alpha\alpha} & V_{\alpha\beta} \\ V_{\beta\alpha} & V_{\beta\beta} + V_{\beta\gamma}G_\gamma(E) V_{\gamma\beta} \end{array}\right)\,.
\end{equation}
Eq.~(\ref{eq:Tb_Tr}) gives the  resonant part of the on-shell T-matrix element for elastic scattering in the $\alpha$-channel,
\begin{equation}
T_{\alpha\alpha}^r  = \braket{\phi_P^{\alpha-}|H_{PQ}\frac{1}{E-H_{QQ}}H_{QP}|\Psi^+}\,.
\end{equation}
The explicit form of the operator in this expression is
\begin{equation}
H_{PQ}\frac{1}{E-H_{QQ}}H_{QP} = \left( \begin{array}{cc} 0 & 0 \\ 0 & V_{\beta\gamma}G_\gamma V_{\gamma\beta} \end{array}\right)\,,
\end{equation} 
so
\begin{equation}
T_{\alpha\alpha}^r = \braket{\beta_{\mathrm{BG}}^-|V_{\beta\gamma}G_\gamma V_{\gamma\beta}|\beta^+}\,.
\end{equation}
Because the resonance is due to a single isolated bound state in the $\gamma$-channel  we may approximate
\begin{equation}
G_\gamma(E)  \approx \frac{\ket{\phi_\gamma}\bra{\phi_\gamma}}{E-\epsilon_\gamma}\,,
\end{equation}
and
\begin{equation}
T_{\alpha\alpha}^r =   \frac{ 
\braket{\beta_{\mathrm{BG}}^-|V_{\beta\gamma}|\phi_\gamma}
\braket{\phi_\gamma|V_{\gamma\beta}|\beta^+} }{E-\epsilon_\gamma}\,.\label{eq:anotherTr}
\end{equation}

Eq.~(\ref{eq:psi_implicit}) becomes
\begin{align}
\ket{\alpha^+} & =  \ket{\alpha_{\mathrm{BG}}^+} \\
\ket{\beta^+} & = \ket{\beta_{\mathrm{BG}}^+} + \frac{\mathcal{G}^+_{\beta(\alpha+)} V_{\beta\gamma}\ket{\phi_\gamma}\braket{\phi_\gamma|V_{\gamma\beta}|\beta^+}}{E-\epsilon_\gamma}\,.\label{eq:beta_implicit} 
\end{align}
To find the matrix element $\braket{\phi_\gamma|V_{\gamma\beta}\mathcal{G}^+_{\beta(\alpha+)}V_{\beta\gamma}|\phi_\gamma}$, we multiply Eq.~(\ref{eq:beta_implicit}) on the left by $\bra{\phi_\gamma}V_{\gamma\beta}$, and isolate
\begin{equation}
\braket{\phi_\gamma|V_{\gamma\beta}|\beta^+} = \frac{ \braket{ \phi_\gamma | V_{\beta\gamma} |\beta_{\mathrm{BG}}^+} }
{1-\frac{\braket{\phi_\gamma|V_{\gamma\beta}\mathcal{G}^+_{\beta(\alpha+)}V_{\beta\gamma}|\phi_\gamma}}{E-\epsilon_\gamma}}\,.
\end{equation}
Substituting this result into Eq.~(\ref{eq:anotherTr}), we arrive at
\begin{equation}
T_{\alpha\alpha}^r = \frac{
\braket{ \beta_{\mathrm{BG}}^- | V_{\beta\gamma} | \phi_\gamma }
\braket{ \phi_\gamma | V_{\beta\gamma} |\beta_{\mathrm{BG}}^+ } 
}
{ E-\epsilon_\gamma
-\braket{ \phi_\gamma|V_{\gamma\beta}\mathcal{G}^+_{\beta(\alpha+)}V_{\beta\gamma}|\phi_\gamma} 
}\label{eq:Tr1}\,.
\end{equation}

In order to evaluate the matrix element $\braket{ \phi_\gamma|V_{\gamma\beta}\mathcal{G}^+_{\beta(\alpha+)}V_{\beta\gamma}|\phi_\gamma}$ appearing in the denominator of Eq.~(\ref{eq:Tr1}), we need to expand the operator $\mathcal{G}^+_{\beta(\alpha+)}$  in a basis of eigenstates for the operator $H_{\beta\beta}+V_{\beta\alpha}G^+_{\alpha}V_{\alpha\beta}$, which is unknown.  However, since $\mathcal{G}^+_{\beta(\alpha+)}$ is the $\beta\beta$-element of $\frac{1}{E^+-H_{PP}}$, we can use an orthonormal basis of eigenstates for $E-H_{PP}$, given by Eqs.~(\ref{eq:phiPA+2})-(\ref{eq:phiPB+2}), to represent
\begin{equation}
\mathcal{G}^+_{\beta(\alpha+)}(E) 
= \int_0^\infty dE'\,\frac{ \ket{\beta^+_{\mathrm{BG}}(E')}\bra{\beta^+_{\mathrm{BG}}(E')} }{E^+-E'}
+
\frac{\ket{\beta^{\prime+}_{\mathrm{BG}}(E')}\bra{\beta^{\prime+}_{\mathrm{BG}}(E')} }
{E^+-E'}\,.\label{eq:Gba+_Eexpansion}
\end{equation}
Substituting this expansion into the matrix element $\braket{ \phi_\gamma|V_{\gamma\beta}\mathcal{G}^+_{\beta(\alpha+)}(E)V_{\beta\gamma}|\phi_\gamma}$ in Eq.~(\ref{eq:Tr1}), we obtain
\begin{align}
{\mathrm{Re}}\,\braket{ \phi_\gamma|V_{\gamma\beta}\mathcal{G}^+_{\beta(\alpha+)}(E)V_{\beta\gamma}|\phi_\gamma}
& = {\mathcal{P}}\int_0^\infty \frac{ |\braket{\phi_\gamma| V_{\gamma\beta}|\beta_{\mathrm{BG}}^+(E')}|^2 + 
|\braket{\phi_\gamma| V_{\gamma\beta}|\beta^{\prime+}_{\mathrm{BG}}(E')}|^2
}{E-E'}\nonumber\\
 &\equiv\Delta(E)\,,\label{eq:defDelta}
\end{align}
where ${\mathcal{P}}$ indicates the Cauchy principal value integral.  With the use of Eq.~(\ref{eq:deltafunction}), we obtain
\begin{align}
\mathrm{Im}\,\braket{ \phi_\gamma|V_{\gamma\beta}\mathcal{G}^+_{\beta(\alpha+)}(E)V_{\beta\gamma}|\phi_\gamma}
& =-\pi\left( \left| \braket{\phi_\gamma| V_{\gamma\beta}|\beta_{\mathrm{BG}}^+(E)}  \right|^2+\left| \braket{\phi_\gamma| V_{\gamma\beta}|\beta^{\prime+}_{\mathrm{BG}}(E)}\right|^2\right)\,\nonumber\\
& \equiv -\frac{\Gamma(E)}{2}\label{eq:defGamma}\,.
\end{align}
Using this definition, Eq.~(\ref{eq:defDelta}), and the definitions
\begin{align}
\Gamma_{\alpha\alpha}(E) & \equiv 2\pi \braket{\beta_{\mathrm{BG}}^-(E)|V_{\beta\gamma}|\phi_\gamma}\braket{\phi_\gamma| V_{\gamma\beta}|\beta_{\mathrm{BG}}^+(E)}\\
\Gamma_{\beta\beta}(E) & \equiv  2\pi \braket{\beta_{\mathrm{BG}}^{\prime-}(E)|V_{\beta\gamma}|\phi_\gamma}\braket{\phi_\gamma| V_{\gamma\beta}|\beta^{\prime+}_{\mathrm{BG}}(E)}\,,\label{eq:defGammaBB}
\end{align}
we obtain
\begin{equation}
T_{\alpha\alpha}^r(E) = \frac{\Gamma_{\alpha\alpha}(E)}
{
2\pi\left( E-\epsilon_\gamma-\Delta(E)+\frac{i\Gamma(E)}{2} \right)
}\,.\label{eq:finalTraa_scattering}
\end{equation}
One may confirm that the definitions in Eqs.~(\ref{eq:defDelta})-(\ref{eq:defGammaBB}) agree with those in Eqs.~(\ref{eq:Deltadefn}), (\ref{eq:Gammadefn}) and (\ref{eq:GammaPPdefn}).
As discussed in Sec.~\ref{sec:feshbach}, $\Gamma(E)=|\Gamma_{\alpha\alpha}(E)|+|\Gamma_{\beta\beta}(E)|$.
As the coupling amplitudes $\Omega_1$, $\Omega_2$ approach zero, $\mathcal{G}^+_{\alpha(\beta+)}\to G^+_\alpha$, $\mathcal{G}^+_{\beta(\alpha+)}\to G^+_\beta$. From Eqs.~(\ref{eq:phiA+andA0}) and (\ref{eq:phiPA+2}) we find that $\ket{\beta^+_{\mathrm{BG}}}={O}(\Omega_1)$, while  from Eqs.~(\ref{eq:phiB+andB0}) and (\ref{eq:phiPB+2}), we obtain $\ket{\beta^{\prime+}_{\mathrm{BG}}}={O}(1)$.   Hence $\Delta(E)={O}(\Omega_2^2)$, $\Gamma_{\alpha\alpha}={O}(\Omega_1^2\Omega_2^2)$, $\Gamma_{\beta\beta}={O}(\Omega_2^2)$ and $\Gamma={O}(\Omega_2^2)$.


\section{Intermediate bound state}
\label{sec:bound}

In this section, we consider the problem in which the intermediate state is a bound state of a closed channel.  The system is illustrated by Fig.~\ref{fig}b.   As in Section~\ref{sec:scattering}, we denote the scattering state solution in the full Hilbert space by $\ket{\Psi^+(E)}$ and define projection operators $P_\alpha$, $P_\beta$ and $P_\gamma$.  In contrast to Sec.~\ref{sec:scattering}, the projection of $\ket{\Psi^+}$ onto the $\beta$-channel is a bound state $P_\beta\ket{\Psi^+}=\ket{\beta}$.
 Our system is then described by the coupled equations,
\begin{align}
(E-H_{\alpha\alpha})\ket{\alpha^+} & = V_{\alpha\beta}\ket{\beta}\label{eq:alphachannel2}\\
(E-H_{\beta\beta})\ket{\beta} & = V_{\beta\alpha}\ket{\alpha^+}+V_{\beta\gamma}\ket{\gamma}\label{eq:betachannel2}\\
(E-H_{\gamma\gamma})\ket{\gamma} & = V_{\gamma\beta}\ket{\beta}\label{eq:gammachannel2}\,.
\end{align}
The bound states $\ket{\phi_\beta}$ and $\ket{\phi_\gamma}$ in the $\beta$- and $\gamma$- channels satisfy the equations $(\epsilon_\beta-H_{\beta\beta})\ket{\beta_0}=0$ and $(\epsilon_\gamma-H_{\gamma\gamma})\ket{\gamma_0}=0$.  We assume that these states are well-separated from other energy eigenstates in their respective channels, both bound and continuum.  Note that $\epsilon_\beta$ and $\epsilon_\gamma$ do not need to be well-separated from each other.

Equations~(\ref{eq:alphachannel2})-(\ref{eq:gammachannel2}) can be solved as described in Sec.~\ref{sec:feshbach}, with the $\alpha$-channel taken as the ${\mathsf{P}}$-channel and the $\beta$- and $\gamma$-channels together forming the ${\mathsf{Q}}$-channel, and the ${\mathsf{Q}}$-channel may be diagonalized as described in Ref.~\cite{feshbach:book} (p.~157). However, it is simpler to apply the technique described in Sec.~\ref{sec:feshbach} twice, first to eliminate the closed $\gamma$-channel, reducing the problem to the $\alpha$- and $\beta$- channels, and then to eliminate the closed $\beta$ channel, which yields an effective equation for the $\alpha$-channel.

Inverting Eq.~(\ref{eq:gammachannel2}) by acting on the left with $G_\gamma$  and substituting the result into Eq.~(\ref{eq:betachannel2}) gives
\begin{equation}
(E-H_{\beta\beta}-V_{\beta\gamma}G_\gamma V_{\gamma\beta})\ket{\beta} = V_{\beta\alpha}\ket{\alpha^+}\,.\label{eq:reducedbetachannel2}
\end{equation}
This eliminates the $\gamma$-channel from the problem and reduces the three-state problem to an effective two-channel problem.  Applying $\mathcal{G}_{\beta(\gamma)}$ to Eq.~(\ref{eq:reducedbetachannel2}) and substituting the result into Eq.~(\ref{eq:alphachannel2}) gives
\begin{equation}
(E-H_{\alpha\alpha}-V_{\alpha\beta}\mathcal{G}_{\beta(\gamma)}V_{\beta\alpha})\ket{\alpha^+} = 0\,.\label{eq:effectiveschrodingeralpha2}
\end{equation}
The effective potential experienced in the $\alpha$-channel is therefore
\begin{equation}
V_{\mathrm{eff}}(E) = V_{\alpha\alpha} +V_{\alpha\beta}\mathcal{G}_{\beta(\gamma)}(E)V_{\beta\alpha}\,.\label{eq:Veff2}
\end{equation}
Rearranging Eq.~(\ref{eq:effectiveschrodingeralpha2})  and multiplying on the left by $G_\alpha^+$ gives
\begin{equation}
\ket{\alpha^+} =  \ket{\phi_{\alpha}^+} + G_\alpha^+V_{\alpha\beta}\mathcal{G}_{\beta(\gamma)}V_{\beta\alpha}\ket{\alpha^+}\,,\label{eq:alpha+implicit}
\end{equation}
where $\ket{\phi_\alpha^+}$ is, as in Sec.~\ref{sec:scattering}, a scattering state solution for the uncoupled channel.  The incoming-wave stationary state $\ket{\phi_\alpha^-}$ is related to the free state $\ket{\alpha_0}$ by  
\begin{equation}
\bra{\phi_\alpha^-} =  \bra{\alpha_0} + \bra{\alpha_0}V_{\alpha\alpha}G_\alpha^+ \label{eq:phiA-}\,.
\end{equation}

The on-shell T-matrix element for elastic scattering in the $\alpha$ channel is
\begin{align}
T_{\alpha\alpha} & =  \braket{\alpha_0|V_{\mathrm{eff}}|\alpha^+}\nonumber\\
                           & =  \braket{\alpha_0|V_{\alpha\alpha}|\phi_\alpha^+} + \braket{\phi_\alpha^-|V_{\alpha\beta}\mathcal{G}_{\beta(\gamma)}V_{\beta\alpha}|\alpha^+} \nonumber \\
                           & \equiv  T_{\alpha\alpha}^b + T_{\alpha\alpha}^r\,.
\end{align}                           
Using the isolated state approximation for $\ket{\phi_\beta}$, we represent the Green's operator $\mathcal{G}_{\beta(\gamma)}$ by
\begin{equation}
\mathcal{G}_{\beta(\gamma)}  \approx  \frac{\ket{\phi_\beta}\bra{\phi_\beta}}{E-\epsilon_\beta-\braket{\phi_\beta|V_{\beta\gamma}G_\gamma V_{\gamma\beta}|\phi_\beta}}\,,\label{eq:Gbetagamma}
\end{equation}
to obtain
\begin{equation}
T_{\alpha\alpha}^r = \frac{\braket{\phi_\alpha^-|V_{\alpha\beta}|\phi_\beta}\braket{\phi_\beta|V_{\beta\alpha}|\alpha+}}{E-\epsilon_\beta-\braket{\phi_\beta|V_{\beta\gamma}G_\gamma V_{\gamma\beta}|\phi_\beta}}\,.\label{eq:Tr2}
\end{equation}
The matrix elements $\braket{\phi_\beta|V_{\beta\alpha}|\alpha^+}$ can be evaluated by multiplying Eq.~(\ref{eq:alpha+implicit}) on the left by $\bra{\phi_\beta}V_{\beta\alpha}$:
\begin{equation}
\braket{\phi_\beta|V_{\beta\alpha}|\alpha^+} = \frac{\braket{\phi_\beta|V_{\beta\alpha}|\phi_\alpha^+}}{1 - \frac{\braket{\phi_\beta|V_{\beta\alpha}G_\alpha^+V_{\alpha\beta}|\phi_\beta}}{E-\epsilon_\beta-\braket{\phi_\beta|V_{\beta\gamma}G_\gamma V_{\gamma\beta}|\phi_\beta}}}\,.
\end{equation}
Substituting this result into Eq.~(\ref{eq:Tr2}) yields
\begin{equation}
T_{\alpha\alpha}^r = 
\frac{\braket{\phi_\alpha^-|V_{\alpha\beta}|\phi_\beta}\braket{\phi_\beta|V_{\beta\alpha}|\phi_\alpha^+}}{E-\epsilon_\beta-\braket{\phi_\beta|V_{\beta\alpha}G_\alpha^+V_{\alpha\beta}|\phi_\beta}-\braket{\phi_\beta|V_{\beta\gamma}G_\gamma V_{\gamma\beta}|\phi_\beta}}\,.\label{eq:Tr3}
\end{equation}
Using the isolated state approximation for $\ket{\phi_\gamma}$, the Green's operator $G_\gamma$ can be written as
\begin{equation}
G_\gamma(E) \approx \frac{\ket{\phi_\gamma}\bra{\phi_\gamma}}{E-\epsilon_\gamma}\,,\label{eq:Ggamma}
\end{equation}
and hence
\begin{equation}
\braket{\phi_\beta|V_{\beta\gamma}G_\gamma V_{\gamma\beta}|\phi_\beta} = \frac{|\braket{\phi_\beta|V_{\beta\gamma}|\phi_\gamma}|^2}{E-\epsilon_\gamma} \equiv \frac{A
}{E-\epsilon_\gamma}\,.\label{eq:Vbgb}
\end{equation}
The states $\ket{\phi_\alpha^+(E')}$ form an orthonormal basis for the $\alpha$-channel so we may expand $G_\alpha^+$ as
\begin{equation}
G_\alpha^+(E)  =  \int_0^\infty dE'\frac{\ket{\phi_\alpha^+(E')}\bra{\phi_\alpha^+(E')}}{E^+-E'}\,,
\end{equation}
to obtain
\begin{equation}
 {\mathrm{Re}}\,\braket{\phi_\beta|V_{\beta\alpha}G_\alpha^+(E)V_{\alpha\beta}|\phi_\beta}
= \mathcal{P}\int_0^\infty dE' \frac{|\braket{\phi_\beta|V_{\beta\alpha}|\phi_\alpha^+(E')}|^2}{E-E'} \equiv \Delta(E)\,,\label{eq:Re}
\end{equation}
where ${\mathcal{P}}$ indicates the Cauchy principal value integral, and
\begin{equation}
{\mathrm{Im}}\,\braket{\phi_\beta|V_{\beta\alpha}G_\alpha^+(E)V_{\alpha\beta}|\phi_\beta} = -\pi|\braket{\phi_\beta|V_{\beta\alpha}|\phi_\alpha^+(E)}|^2 \equiv -\frac{\Gamma(E)}{2}\,.\label{eq:Im}
\end{equation}
Inserting Eqs.~(\ref{eq:Vbgb}), (\ref{eq:Re}) and (\ref{eq:Im}) into Eq.~(\ref{eq:Tr3}) we have
\begin{equation}
T_{\alpha\alpha}^r  = \frac{
\braket{\phi_\alpha^-(E)|V_{
\alpha\beta}|\phi_\beta}
\braket{\phi_\beta|V_{\beta\alpha}|\phi_\alpha^+(E)}
}{2\pi\left(E-\epsilon_\beta - \Delta(E)+\frac{i\Gamma(E)}{2}-\frac{A}{E-\epsilon_\gamma}\right)}\,.
\label{eq:Tr4}
\end{equation}
From Eq.~(\ref{eq:factoring1}), 
\begin{equation}
\bra{\phi_\alpha^-} = (1-2\pi i \braket{\alpha_0|V_{\alpha\alpha}|\phi_\alpha^+})\bra{\phi_\alpha^+}=s^b_\alpha\bra{\phi_\alpha^+}\,,
\end{equation}
and with Eq.~(\ref{eq:Tr4}) we obtain
\begin{equation}
T^r_{\alpha\alpha} = \frac{e^{2i\delta_b}\Gamma}{E-\epsilon_\beta-\Delta+\frac{i\Gamma}{2}-\frac{A}{E-\epsilon_\gamma}}\,.\label{eq:finalTr_bound}
\end{equation}
The S-matrix element for the $\alpha$-channel $s_\alpha$ is defined in Eq.~(\ref{eq:defslittles}). From the discussion in the last paragraph of Sec.~\ref{sec:feshbach}, it satisfies the equations $s_\alpha=e^{2i(\delta_b+\delta_r)}=s^b_\alpha-2\pi iT^r_{\alpha\alpha}$, therefore
\begin{equation}
e^{2i\delta_r} = 1-2\pi ie^{-2i\delta_b}T^r_{\alpha\alpha}
= \frac{(E-\epsilon_\gamma)(E-\epsilon_\beta-\Delta-\frac{i\Gamma}{2})-A}{(E-\epsilon_\gamma)(E-\epsilon_\beta-\Delta+\frac{i\Gamma}{2})-A}\,,
\label{eq:deltar}
\end{equation}
and the resonant phase shift is 
\begin{equation}
\delta_r = \arg \left[(E-\epsilon_\gamma)\left(E-\epsilon_\beta-\Delta-\frac{i\Gamma}{2}\right)-A\right]\,.\label{eq:deltar2}
\end{equation}

Feshbach resonances correspond to poles of the $S$-matrix, and therefore of $T_{\alpha\alpha}^r$, which, if $A\neq0$, occur when
\begin{equation}
(E-\epsilon_\gamma)\left(E-\epsilon_\beta-\Delta(E)+\frac{i\Gamma(E)}{2}\right) - A = 0\,.\label{eq:almostquadratic}
\end{equation} 
Denoting the roots of this equation by $E_1$ and $E_2$, we see that in the limit of $A\to0$,  $E_1\to\epsilon_\gamma$ and $E_2-\epsilon_\beta-\Delta(E_2)+\frac{i\Gamma(E_2)}{2}\to0$.  For $A=0$ the $\alpha$ and $\beta$ channels decouple from the $\gamma$ channel and the pole at $E_1$ is not present.  The quantities $\Delta(E)$ and $\Gamma(E)$ usually vary slowly with $E$ so there should be two resonances, one associated with the bound state $\ket{\phi_\beta}$ and one with $\ket{\phi_\gamma}$.  If we assume that $\Delta(E)$ and $\Gamma(E)$ are independent of energy, then, to first order in $A$, the poles occur at
\begin{align}
E_1 & =  \epsilon_\gamma -\frac{2A}{\epsilon_\beta+\Delta-\frac{i\Gamma}{2}-\epsilon_\gamma}\nonumber\\
E_2 & =  \epsilon_\beta+\Delta-\frac{i\Gamma}{2} +\frac{2A}{\epsilon_\beta+\Delta-\frac{i\Gamma}{2}-\epsilon_\gamma}\,.\label{eq:quadraticroots}
\end{align}
The poles move in opposite directions with increasing $A$.  In particular, the pole at $E_1$ moves away from the real axis, which results in a non-zero resonance width.  In the limit of $A\to\infty$,
\begin{equation}
E_{1,2}=\frac{\epsilon_\gamma+\epsilon_\beta+\Delta-\frac{i\Gamma}{2}}{2}\pm A^{1/2}(1+O(A^{-1}))\;,
\end{equation} 
where the $+$ root is $E_1$ if $\epsilon_\gamma>\epsilon_\beta+\Delta$ and $E_2$ if $\epsilon_\beta+\Delta> \epsilon_\gamma$.  

The coupling between channels $\alpha$ and $\beta$ causes a displacement in the position of the $\ket{\phi_\beta}$ resonance to $\epsilon_\beta+\Delta$ when $A=0$, hence degeneracy of the resonances occurs when $\epsilon_\gamma=\epsilon_\beta+\Delta$.  The resonance poles are then located at $E_{1,2}=\epsilon_\gamma-i\Gamma/4\pm(A-(\Gamma/4)^2)^{1/2}$.  If $4A^{1/2}<\Gamma$ there are two resonances at the same energy with different widths, if $4A^{1/2}>\Gamma$ there are two resonances at different energies, both with width $\Gamma/4$, and when $\Gamma=4A^{1/2}$ there is a single pole of order two.  

\begin{figure}
\scalebox{0.9}{\includegraphics{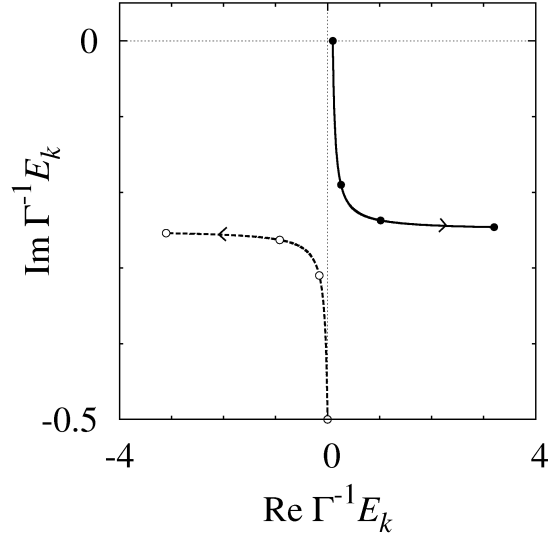}}
\caption{\label{fig:quadraticroots}Trajectories of the roots of $(E-\epsilon_\gamma)(E-\Delta-\epsilon_\beta+i\frac{\Gamma}{2})-A$ in the complex $E$-plane as $A$ increases.    The calculations are performed for $(\epsilon_\beta+\Delta)/\Gamma=0$, $\epsilon_\gamma/\Gamma=0.1$, and a finely-spaced grid of $A/\Gamma$ values.  Symbols indicate the points on each trajectory corresponding to $A/\Gamma=0$, $0.1$, $1$ and $10$.} 
\end{figure}

Fig.~\ref{fig:quadraticroots} illustrates motion of the resonance poles in the complex energy plane as $A$ varies for a system with constant $\Delta$ and $\Gamma$, for a nondegenerate case. 

The left-hand side of Eq.~(\ref{eq:deltar}) can be factored as
\begin{equation}
\left(E-{\mathcal{E}}_1(E)+\frac{i\Gamma_1(E)}{2}\right)\left(E-{\mathcal{E}}_2(E)+\frac{i\Gamma_{2}(E)}{2}\right)\,,
\label{eq:factorization}
\end{equation}
where the functions ${\mathcal{E}}_1(E)$, ${\mathcal{E}}_2(E)$, $\Gamma_1(E)$ and $\Gamma_2(E)$ are real-valued for real $E$ and satisfy
\begin{align}
\mathcal{E}_1(E) + \mathcal{E}_2(E) -i\left(\frac{\Gamma_1(E)+\Gamma_2(E)}{2}\right) 
&=\epsilon_\gamma+\epsilon_\beta+\Delta(E)-i\frac{\Gamma(E)}{2}\nonumber\\
\mathcal{E}_1(E)\mathcal{E}_2(E)-\frac{\Gamma_1(E)\Gamma_2(E)}{4}-i\left(\frac{\mathcal{E}_1(E)\Gamma_2(E) +\mathcal{E}_2(E)\Gamma_1(E)}{2}\right)  & = \epsilon_\gamma(\epsilon_\beta+\Delta(E))-A-i\frac{\epsilon_\gamma \Gamma(E)}{2}\,.
\end{align}
Each pole of $T^r_{\alpha\alpha}$ corresponds to a zero of one of the factors and the resonant phase shift from  Eq.~(\ref{eq:deltar2}) can be expressed as the sum of phase shifts for each resonance,
\begin{equation}
\delta_r(E) = \delta^r_1(E)+\delta^r_2(E) = \arg\left(E-{\mathcal{E}}_1(E)-\frac{i\Gamma_1(E)}{2}\right) + \arg\left(E-{\mathcal{E}}_2(E)-\frac{i\Gamma_2(E)}{2}\right)\,.
\end{equation}

As the coupling amplitudes $\Omega_1$ and $\Omega_2$ vanish, $A={O}(\Omega_2^2)$, $\Gamma={O}(\Omega_1^2)$ and $\Delta={O}(\Omega_1^2)$.  If $\Omega_1$ and $\Omega_2$ are small compared to the distance between the zeroth order resonance positions, $\epsilon_\beta+\Delta-\epsilon_\gamma$, the energy shift for the resonance due to $\ket{\phi_\gamma}$ is ${\mathrm{Re}}(E_1)-\epsilon_\gamma={O}(\Omega_2^2)$, and the resonance width is ${\mathrm{Im}}(E_1)/2={O}(\Omega_1^2\Omega_2^2)$.  The resonance due to $\ket{\phi_\beta}$ has an energy shift ${\mathrm{Re}}(E_2)-\epsilon_\beta={O}(\Omega_1^2)$ and width  ${\mathrm{Im}}(E_2)/2={O}(\Omega_1^2)$. 
  
  
\section{Multiple intermediate bound states}
\label{sec:multiple}

Here, we  generalize the results of the previous section to the problem with several intermediate bound states, each in a different closed channel, coupled in a chain to the resonant bound state.  We use the same notation as before, except that it is now convenient to label channel indices with numbers rather than letters, and to place the open channel last in the indexing scheme.  Our system is described by the coupled equations,
\begin{align}
(E-H_{11})\ket{1} & = V_{12}\ket{2}\nonumber\\
(E-H_{22})\ket{2} & = V_{21}\ket{1} +V_{32}\ket{3}\nonumber\\
& \vdots  \nonumber\\
(E-H_{nn})\ket{n^+} & = V_{n-1,n}\ket{n-1}\,,\label{eq:system}
\end{align}
where channel $n$ is open and contains the incident flux and channels 1 through $n-1$ are closed.  We assume that each closed channel $k$ contains a bound state $\ket{\phi_k}$ satisfying $(E-H_{kk})\ket{\phi_k}=0$, and that each such state is well-separated from other states in its channel.  We adopt the procedure of the previous section to repeatedly remove the last closed channel.  After eliminating channels 1 and 2, we obtain the following equation for channel 3:
\begin{equation}
(E-H_{33} - V_{32}\mathcal{G}_{2(1)}V_{23})\ket{3}  = V_{34}\ket{4}\,.\label{eq:channel3}
\end{equation}

For the more complicated Green's operators that appear in solving this system, such as that for the pseudo-Hamiltonian on the left-hand side of Eq.~(\ref{eq:channel3}), we extend the notation defined in Sec.~\ref{sec:notation} with a recursive definition.  If $\sigma_k$ with $k=1$ to $n$ are channel indices,
\begin{equation}
\mathcal{G}_{\sigma_n(\sigma_{n-1}(\cdots(\sigma_1)\cdots))}(E) \equiv \frac{1}{E-H_{\sigma_n\sigma_n} - V_{\sigma_n,\sigma_{n-1}}\mathcal{G}_{\sigma_{n-1}(\sigma_{n-2}(\cdots(\sigma_1)\cdots))}(E)V_{\sigma_{n-1}\sigma_n}} \,.
\end{equation}
The inverse of the operator in Eq.~(\ref{eq:channel3}) is denoted by $\mathcal{G}_{3(2(1))}(E)$.  Repeating this process we obtain for channel $n$ an effective Schr\"{o}dinger equation
\begin{equation}
(E-H_{nn}-V_{n,n-1}\mathcal{G}_{n-1(n-2(\cdots(1)\cdots))}V_{n-1,n})\ket{n^+} = 0\,,
\end{equation}
with the effective potential determined by 
\begin{equation}
V_{\mathrm{eff}}(E)  =  V_{nn} + V_{n,n-1}\mathcal{G}_{n-1(n-2(\cdots(1)\cdots))}(E)V_{n-1,n}\,,
\end{equation}
where the operator $\mathcal{G}_{n-1(n-2(\cdots(1)\cdots))}(E)$ is
\begin{multline}
\mathcal{G}_{n-1(n-2(\cdots(1)\cdots))}(E)  \\ 
=\cfrac{1}{E-H_{n-1,n-1}-V_{n-1,n-2}\cfrac{1}{E-H_{n-2,n-2}-V_{n-2,n-3}\cfrac{1}{\cfrac{\vdots}{E-H_{22}-V_{21}\cfrac{1}{E-H_{11}}V_{12}}}V_{n-3,n-2}}V_{n-2,n-1}}\,.
\end{multline}
Using Eq.~(\ref{eq:Ggamma}) repeatedly we obtain,  for $k=1,\ldots,n-1$,
 \begin{equation}
\mathcal{G}_{k(k-1(\cdots))}(E) \approx \ket{\phi_k}\bra{\phi_k} \frac{P_{k-1}(E)}{P_k(E)}\,,
\end{equation}
where $P_k(E)$ are polynomials in $E$ satisfying the recursion relation
\begin{align}
P_0(E) &= 1\nonumber\\
P_1(E) &= E-\epsilon_1\nonumber\\
P_k(E) & =  (E-\epsilon_k)P_{k-1}(E) - A_kP_{k-2}(E)\,,\qquad k\geq2\,,\label{eq:recursion}
\end{align}
where
\begin{align}
A_k & \equiv  |\braket{\phi_k|V_{k,k-1}|\phi_{k-1}}|^2\,.
\end{align}
This gives the implicit equation for $\ket{n^+}$,
\begin{equation}
\ket{n^+} = \ket{\phi_n^+} + G_n V_{n,n-1}\ket{\phi_{n-1}}\braket{\phi_{n-1}|V_{n-1,n}|n^+}\frac{P_{n-2}}{P_{n-1}}\,,\label{eq:n_implicit}
\end{equation}
from which we can obtain the resonant contribution to the T-matrix element for elastic scattering in the open channel,
\begin{equation}
T_{nn}^r =  \braket{\phi_n^-|V_{n,n-1}|\phi_{n-1}}\braket{\phi_{n-1}|V_{n,n-1}|n^+}\frac{P_{n-2}}{P_{n-1}}\,.\label{eq:Tr_multiple1}
\end{equation}
We evaluate the matrix element $\braket{\phi_{n-1}|V_{n,n-1}|n^+}$ by multiplying Eq.~(\ref{eq:n_implicit}) on the left by $\bra{\phi_{n-1}}V_{n-1,n}$:
\begin{equation}
\braket{\phi_{n-1}|V_{n,n-1}|n^+} = \frac{\braket{\phi_{n-1}|V_{n-1,n}|\phi_n^+}}{1-\braket{\phi_{n-1}|V_{n-1,n} G_n V_{n,n-1}|\phi_{n-1}}\frac{P_{n-2}}{P_{n-1}}}\,.
\end{equation}
Using this result, Eq.~(\ref{eq:Tr_multiple1}) yields
\begin{align}
T_{nn}^r(E) & =  \frac{\braket{\phi_n^-|V_{n,n-1}|\phi_{n-1}}\braket{\phi_{n-1}|V_{n-1,n}|\phi_n^+}P_{n-2}(E)}{P_{n-1}(E)-\braket{\phi_{n-1}|V_{n-1,n}G_n(E)V_{n,n-1}|\phi_{n-1}}P_{n-2}(E)} \nonumber\\
& =  \frac{\Gamma(E)P_{n-2}(E)}{2\pi Q_{n-1}(E)}\,,\label{eq:Tr_multiple2}
\end{align}
with $Q_{n-1}(E)$ defined by  
\begin{equation}
Q_{n-1}(E) = P_{n-1}(E)-\left(\Delta(E)-\frac{i\Gamma(E)}{2}\right)P_{n-2}(E)\,,\label{eq:Qdefn}
\end{equation}
where in these last two equations $\Gamma(E)$ and $\Delta(E)$ are as given by Eqs.~(\ref{eq:Re}) and (\ref{eq:Im}) but with $\alpha$ replaced by $n$ and $\beta$ by $n-1$.

Resonances are associated with the roots of the equation $Q_{n-1}(E)=0$, which correspond to poles of the $S$-matrix.  The root which approaches $\epsilon_k$ (or $\epsilon_{n-1}+\Delta(\epsilon_n-1)$, when $k=n-1$) as the coupling strengths $A_i$ tend to zero is $E_k$.  The real part of the root gives the resonance energy, and the width is $-2\,{\mathrm{Im}}\,E_k$.  A real root of $Q_{n-1}$, for example when one of the couplings $A_k=0$, must be a root of both $P_{n-1}$ and $P_{n-2}$, and by Eq.~(\ref{eq:Tr_multiple2}) does not give rise to a pole in $T_{nn}^{r}$.  Physically, the resonance width approaches zero as the root approaches the real axis.  A factorization
\begin{equation}
Q_{n-1}(E) = \prod_{k=1}^{n-1}E-\mathcal{E}_k(E)+\frac{i\Gamma_k(E)}{2}
\label{eq:factorization2}
\end{equation}
exists, where $\mathcal{E}_k(E)$ and $\Gamma_k(E)$ are real for real $E$,  $E_k=\mathcal{E}_k(E_k)-\frac{i\Gamma_k(E_k)}{2}$ and
\begin{equation}
\sum_{k=1}^{n-1}\left(\mathcal{E}_k(E)-\frac{i\Gamma_k(E)}{2}\right)=\Delta(E)-\frac{i\Gamma(E)}{2}+\sum_{k=1}^{n-1}\epsilon_k\,.
\label{eq:property}
\end{equation}

\begin{figure}
\scalebox{0.9}{\includegraphics{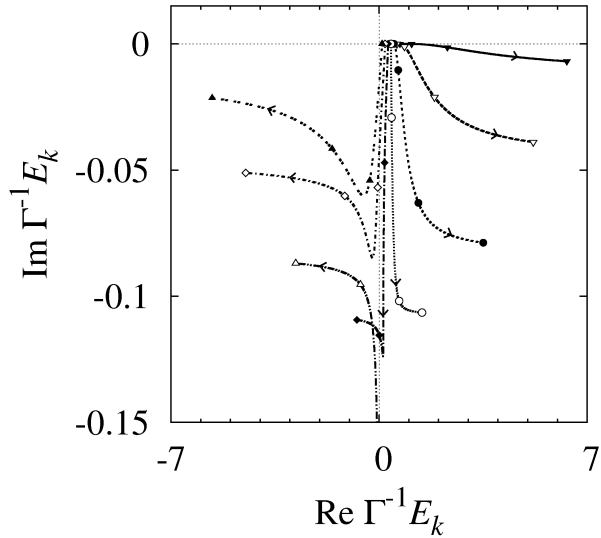}}
\caption{\label{fig:eighthdegreeroots}Trajectories  of the roots of an eighth-degree polynomial $Q_8(E)$ in the complex $E$-plane as $A$ increases.  Calculations are performed with all $A_k=A$,  $(\epsilon_8+\Delta)/\Gamma=0$, $\epsilon_k/\Gamma=0.1(8-k)$ for $k=1,\ldots,7$, on a fine grid of $A/\Gamma$ values.  Symbols indicate the points corresponding to  $A/\Gamma=0$, $0.1$, $1$ and $10$.}
\end{figure}

In order to elucidate the properties of the resonances induced by indirect coupling, we consider the case with constant $\Delta$ and $\Gamma$ and all $A_k$ equal to a constant $A$.  $Q_{n-1}(E)$ is then a polynomial in $E$ and $A$, which may be solved numerically.  $\mathcal{E}_k-i\Gamma_k/2=E_k$ in Eqs.~(\ref{eq:factorization2}) and (\ref{eq:property}) implies that the roots of the polynomial sum to $\Delta-i\Gamma/2+\sum \epsilon_k$.   Fig.~\ref{fig:eighthdegreeroots} illustrates the trajectories of the roots of an eighth-degree polynomial $Q_8(E)$ in the complex energy plane as $A$ increases.  The energies of the bound states $\epsilon_k$ are assumed to be closely and regularly spaced.  For all roots, ${\mathrm{Im}}\,E_k<0$ when $A>0$, and ${\mathrm{Im}}\,E_k$ approaches a constant as $A\to\infty$.   We have solved a range of other examples, varying the ordering of $\epsilon_k$, the regularity and magnitude of their spacing, and the number of bound states $n-1$, and found that this is a generic property.  If all roots lie in the ${\mathrm{Im}}\,E<0$ half-plane, the resonance widths sum to $\Gamma$ and every resonance must have width less than $\Gamma$.  It can be proven that the roots become real only when $A=0$ so each root must remain on one side of the real axis for $A>0$.  From Eq.~(\ref{eq:Qdefn}), a real root of $Q_{n-1}$ must be a root of both $P_{n-1}$ and $P_{n-2}$.  From Eq.~(\ref{eq:recursion}), $AP_{k-2}=(E-\epsilon_k)P_{k-1}-P_{k}$ for $k\geq2$, hence if $A\neq0$, a common root of $P_{n-1}$ and $P_{n-2}$ must be a root of all $P_k$, $k\leq n-1$, including $P_0$.  However, this would imply that $P_0$ has a root, which is not possible since $P_0=1$.  We conclude that in a solution of Eq.~(\ref{eq:recursion}) with $A\neq0$, no two consecutive $P_k$ may have a common root.  As a consequence, there can be no real roots of $Q_{n-1}$ unless $A=0$.  

\begin{figure}
\scalebox{0.9}{\includegraphics{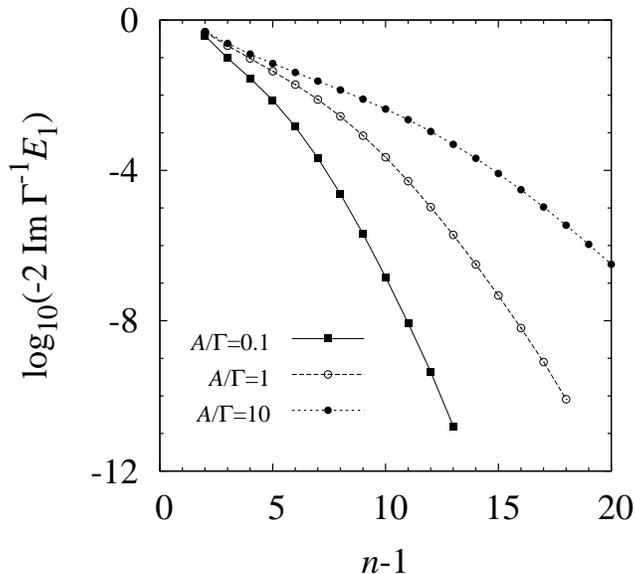}}
\caption{\label{fig:widths}Width of the resonance due to $\ket{\phi_1}$ for a system with $n-1$ bound states calculated with all $A_k=A$, $(\epsilon_{n-1}+\Delta)/\Gamma=0$, $\epsilon_k/\Gamma=0.1(n-1-k)$ for $k=1,\ldots,n-2$. }
\end{figure}

An analysis of the resonance widths shows that the width of the resonance at $E_1$ due to the last bound state $\ket{\phi_1}$ decreases rapidly as the number of intermediate bound states increases (Fig.~\ref{fig:widths}).

\section{Summary}
\label{sec:conc}

We have presented analytical solutions of the Schr\"{o}dinger equation describing multi-channel Feshbach resonances mediated by second- and higher-order couplings, {\textit i.e.} scattering resonances  induced by the interaction with a bound state that is not directly coupled to the initial scattering state. The results obtained represent three distinct problems: (i) resonant scattering induced by a sequence of one continuum-continuum and one continuum-bound couplings; (ii) resonant scattering induced by a sequence of one continuum-bound and one bound-bound couplings; and (iii) resonant scattering induced by a sequence of one continuum-bound and several bound-bound couplings. 

The first problem is a model of electric-field-induced resonances in ultracold collisions of alkali metal atoms \cite{krems:prl,likrems:pra}.  Electric fields couple $s$-wave collision channels with $p$-wave scattering states. The numerical calculations of Refs.~[\onlinecite{krems:prl}] and [\onlinecite{likrems:pra}] showed that the $s$-wave collision cross section may undergo a resonant variation in the presence of electric fields if the corresponding $p$-wave collision channel is coupled resonantly with a $p$-wave bound state. Section~\ref{sec:scattering} of the present paper shows that this is a general phenomenon. Eqs.~(\ref{eq:defDelta}) - (\ref{eq:finalTraa_scattering}) provide a general description of such three-state Feshbach resonances and demonstrate how the width and position of the resonances in the initial scattering state depend on the strengths of the continuum-continuum ($s$-to-$p$) and continuum-bound ($p$-to-$p$) couplings.  Similar resonances may arise in collisions of atoms or molecules with significant magnetic or electric dipole - dipole interactions. The dipole - dipole interactions couple different partial waves of the collision complex and some of the dipolar resonances observed in ultracold collisions of Cr atoms \cite{pavlovic:pra,werner:prl,stuhler:jmo} may have a three-state character.  Eq.~(\ref{eq:finalTraa_scattering}) of the present work provides a general form that can be used to fit the experimental data on three-state Feshbach resonances involving two continuum states and the theory of Sec.~\ref{sec:scattering} may be used for a refined analysis of Feshbach resonances in the Bose-Einstein condensate of Cr atoms \cite{pavlovic:pra,werner:prl,stuhler:jmo}. 

The second and third problems considered in this work are relevant for collisions of complex polyatomic molecules with ultracold atoms or molecules. Polyatomic molecules can potentially be cooled to ultracold temperatures by elastic collisions in a reservoir of ultracold atoms \cite{modugno:science,ryjkov:pra,ostendorf:prl}. The experimental realization of this method may be complicated by naturally occurring Feshbach resonances. The energy spectrum of polyatomic molecules is quite dense and collisions of large molecules with ultracold atoms may lead to long-lived Feshbach resonances that would complicate translational energy exchange and result in sticking of atoms to molecules and the formation of clusters. It is therefore extremely important to understand the mechanisms of Feshbach resonances in collisions of polyatomic molecules with atoms. 

If the molecule is sufficiently large, the atom - molecule scattering state of interest may not be directly coupled to all molecular states in a collision. The atom - molecule interaction potential, however, induces couplings between different states of the molecule and the entire spectrum of molecular states may be coupled to the atom - molecule scattering state through a sequence of one continuum - bound and several bound - bound couplings. The simplest example of this coupling mechanism is a collision system of a structureless atom and a diatomic molecule interacting through the long-range dispersion interaction. The dispersion interaction couples the ground rotational state $N=0$ of the molecule only with the first and second rotationally excited states $N=1$ and $N=2$; however,  the bound states of the atom - molecule complex corresponding asymptotically to $N > 2$  may give rise to Feshbach resonances in collisions of ground-state molecules through a sequence of $N > 2$ -- $N=2$ and $N=2$ -- $N=0$ couplings.  

Section~\ref{sec:bound} presents a general analysis of three-state Feshbach resonances induced by one continuum - bound and one bound - bound couplings. Eq.~(\ref{eq:finalTr_bound}) gives the general form of  the resonant variation of the T-matrix element and Eqs.~(\ref{eq:almostquadratic}) and (\ref{eq:quadraticroots}) show that the two bound states give rise to two resonances.  Fig.~\ref{fig:quadraticroots} illustrates the dependence of the resonance positions and widths on the coupling strength.  Our analysis shows that the scattering amplitude must exhibit two resonances, even if the two bound states are degenerate.  The continuum-bound coupling shifts the resonance energy of the first bound state, and degeneracy occurs when the shifted energy equals the second bound state energy.  For small coupling strengths, the two resonances occur at the same position, while for large coupling strengths, they have the same width.

Section~\ref{sec:multiple} generalizes the results of Sec.~\ref{sec:bound} for systems with several intermediate bound-bound couplings and demonstrates that Feshbach resonances may occur even if the scattering state  is separated from the resonant bound state by a sequence of several indirectly coupled bound states.  The ladder character of the couplings ensures that the scattering amplitude exhibits a pole near the energy of the bound state.  We have shown that the energy dependence of the T-matrix element can be written in a general form given by Eq.~(\ref{eq:Tr_multiple2}). The polynomials $P_k$ in Eq.~(\ref{eq:Tr_multiple2}) depend on the structure of the molecule and the atom - molecule interaction potentials.  They can be evaluated using the recursive procedure described by Eqs.~(\ref{eq:recursion}).  A numerical analysis of the polynomial roots shows that the width of the resonance decreases rapidly as the number of intermediate off-resonant bound states increases.  Resonances induced by high-order couplings should be ubiquitous in collision systems involving complex molecules with multiple degrees of freedom and our expressions and formalism can be used for the analysis of experiments on ultracold collisions of atoms and molecules, chemical reaction dynamics and electron - molecule scattering. 

\begin{acknowledgments}
The work was supported by NSERC of Canada.
\end{acknowledgments}

\end{document}